\documentclass[11pt]{article}
\usepackage{graphicx}
\usepackage[margin=1.25in]{geometry}
\usepackage[usenames,dvipsnames]{color}
\usepackage{url}
\usepackage[normalem]{ulem}
\usepackage[colorlinks = true,
            linkcolor = blue,
            urlcolor  = blue,
            citecolor = blue,
            anchorcolor = blue]{hyperref}

\usepackage{makecell}
\usepackage{hyperref,color}
\usepackage{todonotes}
\usepackage{booktabs}
\usepackage{float}
\usepackage{wrapfig}

\usepackage{appendix}
\usepackage{pdfpages}
\usepackage{hyperref}
\usepackage{geometry}                
\usepackage[parfill]{parskip}    
\usepackage{graphicx, subfigure}
\usepackage{amssymb}
\usepackage{amsmath}
\usepackage{epstopdf}
\usepackage{array}
\usepackage{lineno}
\usepackage{scrextend}

\def\Title#1{\begin{center} {\LARGE #1 } \end{center}}
\def\Author#1{\begin{center}{ \sc #1} \end{center}}
\def\Address#1{\begin{center}{ \it #1} \end{center}}

\newenvironment{Abstract}{\begin{quotation} \begin{center}
                       ABSTRACT
     \end{center}\bigskip  }{\end{quotation}}

\def\Acknowledgements{\bigskip  \bigskip \begin{center} \begin{large}
             \bf ACKNOWLEDGEMENTS \end{large}\end{center}}

\def\Acknowledgements{\bigskip  \bigskip \begin{center} \begin{large}
             \bf ACKNOWLEDGEMENTS \end{large}\end{center}}

\def\beq{\begin{equation}}
\def\eeq#1{\label{#1}\end{equation}}
\def\eeqn{\end{equation}}

\newenvironment{Eqnarray}%
   {\arraycolsep 0.14em\begin{eqnarray}}{\end{eqnarray}}
\def\beqa{\begin{Eqnarray}}
\def\eeqa#1{\label{#1}\end{Eqnarray}}
\def\eeqan{\end{Eqnarray}}

\let\bar=\overbar

\def\lsim{\mathrel{\raise.3ex\hbox{$<$\kern-.75em\lower1ex\hbox{$\sim$}}}}
\def\gsim{\mathrel{\raise.3ex\hbox{$>$\kern-.75em\lower1ex\hbox{$\sim$}}}}

\def\del{\partial}
\def\Dslash{\not{\hbox{\kern-4pt $D$}}}
\def\dslash{\not{\hbox{\kern-2pt $\del$}}}
\def\pslash{\not{\hbox{\kern-2pt $p$}}}
\def\ETmiss{\not{\hbox{\kern-4pt $E$}}_T}

\def\Dlr{\mathrel{\raise1.5ex\hbox{$\leftrightarrow$\kern-1em\lower1.5ex\hbox{$D$}}}}

\def\MSB{{\bar{M \kern -2pt S}}}
\def\msb{{\bar{\scriptsize M \kern -1pt S}}}

\def\drb{{\bar{\scriptsize D \kern -1pt R}}}

\input{defs}

\newcommand\snowmass{\begin{center}\rule[-0.2in]{\hsize}{0.01in}\\\rule{\hsize}{0.01in}\\
\vskip 0.1in Submitted to the  Proceedings of the US Community Study\\ 
on the Future of Particle Physics (Snowmass 2021)\\ 
\rule{\hsize}{0.01in}\\\rule[+0.2in]{\hsize}{0.01in} \end{center}}

\begin{document}


\medskip

\Title{Snowmass White Paper: Probing New Physics with $\mu^+ \mu^- \to bs$ at a Muon Collider}

\bigskip 

\Author{Wolfgang~Altmannshofer, Sri Aditya Gadam, and Stefano Profumo}
\Address{Department of Physics and Santa Cruz Institute for Particle Physics\\ University of California, Santa Cruz, CA 95064, USA}

\medskip\medskip

\begin{Abstract}
\noindent In this white paper for the Snowmass process, we discuss the prospects of probing new physics explanations of the persistent rare $B$ decay anomalies with a muon collider.
If the anomalies are indirect signs of heavy new physics, non-standard rates for $\mu^+ \mu^- \to b s$ production should be observed with high significance at a muon collider with center of mass energy of $\sqrt{s} = 10$~TeV. The forward-backward asymmetry of the $b$-jet provides diagnostics of the chirality structure of the new physics couplings. In the absence of a signal, $\mu^+ \mu^- \to b s$ can indirectly probe new physics scales as large as $86$~TeV. Beam polarization would have an important impact on the new physics sensitivity.
\end{Abstract}

\snowmass

\def\thefootnote{\fnsymbol{footnote}}
\setcounter{footnote}{0}


\section{Introduction} \label{sec:intro}

For many years, experimental results on rare decays of $B$ mesons have shown signals consistent with the violation of lepton flavor universality (LFU). The most significant result is the measurement of the LFU ratio $R_{K^+} = \text{BR}(B^+ \to K^+ \mu^+\mu^-)/\text{BR}(B^+ \to K^+ e^+e^-) = 0.846^{+0.044}_{-0.041} $~\cite{LHCb:2021trn}, which differs from the precise Standard Model (SM) prediction, $R_{K^+}^\text{SM} = 1.00 \pm 0.01$~\cite{Bordone:2016gaq,Isidori:2020acz}, by $3.1\sigma$. 
Measurements of LFU ratios in related decays are all consistently low compared to the SM predictions, albeit with lesser significance~\cite{LHCb:2017avl,LHCb:2019efc,LHCb:2021lvy}. From the theory side, the LFU ratios are very well understood: If the significance of the current discrepancies continues to increase with larger statistics, it would be a clear sign of new physics.

Intriguingly, there are several additional experimental results on rare $B$ decays that are in tension with SM predictions. In particular, the branching ratios of the muonic decays $B \to K \mu^+ \mu^-$, $B \to K^* \mu^+ \mu^-$, $B_s \to \phi \mu^+\mu^-$, and to some extent also $B_s \to \mu^+ \mu^-$ are all low compared to SM predictions~\cite{LHCb:2014cxe, LHCb:2016ykl, LHCb:2021zwz, ATLAS:2018cur, CMS:2019bbr, LHCb:2021vsc}. Moreover, the angular distributions in the $B \to K^* \mu^+ \mu^-$ decays additionally show deviations from SM expectations~\cite{LHCb:2020lmf, LHCb:2020gog}. While the LFU ratios could be explained by new physics either in the electron or muon channels, global fits that combine all relevant experimental data on rare $B$ decays indicate strong preference for new physics in the $b \to s \mu\mu$ transition~\cite{Geng:2021nhg, Altmannshofer:2021qrr, Isidori:2021vtc, Alguero:2021anc, Hurth:2021nsi,  Ciuchini:2021smi}.  
Assuming that the new physics is heavy compared to $B$ mesons, it can be described in a model-independent way by dimension 6 operators of an effective Hamiltonian. The leading candidates for an explanation of the $B$ anomalies are the 4 fermion contact interactions $(\bar s \gamma_\alpha P_L b)(\bar \mu \gamma^\alpha \mu)$ and $(\bar s \gamma_\alpha P_L b)(\bar \mu \gamma^\alpha \gamma_5 \mu)$ with a generic new physics scale of approximately $\Lambda_\text{NP} \sim 35$~TeV for ${\mathcal O}(1)$ couplings~\cite{Altmannshofer:2017yso, DiLuzio:2017chi}. It is thus conceivable that the new physics responsible for the rare $B$ anomalies is outside the direct reach of the LHC.
In such a case, measurements of high-mass di-lepton tails at the LHC are in principle still sensitive to non-standard $b s \to \mu^+ \mu^-$ production induced by the contact interactions~\cite{Greljo:2017vvb}. However, the expected sensitivities at the high-luminosity phase of the LHC and even at a future 100 TeV proton-proton collider are very likely insufficient to model independently test a heavy new physics origin of the rare $B$ anomalies.

As discussed in~\cite{Huang:2021nkl, Huang:2021biu, Asadi:2021gah}, a high-energy muon collider~\cite{Ankenbrandt:1999cta, AlAli:2021let} would be able to fully probe $Z^\prime$ and lepto-quark explanations of the rare $B$ anomalies. A muon collider would also enable to conclusively probe the preferred parameter space if the new physics is parameterized in a model-independent way by contact interactions. Even if the corresponding new physics is too heavy to be produced directly, it reveals itself through non standard contributions to the process $\mu^+ \mu^- \to b s$. We find that with a center of mass energy of $\sqrt{s} = 10$~TeV, the new physics contributions can be observed above background with high significance. Moreover, measuring the forward-backward asymmetry of the $b$-jet gives the opportunity to determine the chirality of the new physics couplings to muons.

In this white paper we summarize the main findings of our study~\cite{inprep}. 
In section~\ref{sec:theory}, we introduce the theoretical framework and present the new physics predictions for the signal cross section and the forward-backward asymmetry. The most important background processes and the collider analysis are discussed in section~\ref{sec:backgrounds}. The sensitivity projections of the muon collider to the new physics contact interactions are given in section~\ref{sec:results}.

\section{Theoretical Framework and Signal Cross Section} \label{sec:theory}

New physics in the rare $b\to s \mu\mu$ decays is conveniently parameterized by an effective Hamiltonian consisting of dimension 6 operators with the associated Wilson coefficients. Taking into account only the new physics operators that are known to give a valid explanation of the anomalies, one can write
\begin{equation}
    \label{eq:EffectiveHamiltonian}
    \mathcal H_\text{eff} =  \mathcal H_\text{eff}^\text{SM} - \frac{4G_F}{\sqrt{2}} V_{tb} V_{ts}^* \frac{\alpha}{4\pi} \Big( C_9 O_9 + C_{10} O_{10}\Big) ~,
\end{equation}
with the relevant 4-fermion contact interactions
\begin{equation}
    \label{eq:OperatorChoice}
    O_9 = \left(\overline{s}\gamma_\alpha P_L b\right)\left(\overline{\mu}\gamma^\alpha \mu\right) ~, \quad O_{10} = \left(\overline{s}\gamma_\alpha P_L b\right)\left(\overline{\mu}\gamma^\alpha \gamma_5\mu\right) ~.
\end{equation}
In the effective Hamiltonian~\eqref{eq:EffectiveHamiltonian} we have included the standard normalization factor containing $V_{tb}V_{ts}^*$ that corresponds to the leading CKM factor in the SM. 
The combinations $O_9 - O_{10}$ and $O_9 + O_{10}$ correspond to left-handed and right-handed muon currents, respectively.
The full set of dimension 6 operators will be considered in~\cite{inprep}. Best fit values for the new physics Wilson coefficients $C_9$ and $C_{10}$ are given for example in~\cite{Geng:2021nhg, Altmannshofer:2021qrr, Isidori:2021vtc, Alguero:2021anc, Hurth:2021nsi, Ciuchini:2021smi}. Those best fit values are given at a renormalization scale appropriate for the decay of $B$ mesons. A typical choice is $\mu = 4.2$~GeV. 

Assuming that the description in terms of effective operators is still valid at the energy scale of a muon collider, i.e. the mass of the new physics degrees of freedom is sufficiently higher than the center of mass energy, the differential $\mu^+ \mu^- \to b s$ cross section can be calculated in a straightforward way using the operators above (for many of our calculations we used \verb|FeynCalc|~\cite{MERTIG1991345,Shtabovenko:2016sxi,Shtabovenko:2020gxv}). Allowing for generic polarizations of the muon and anti-muon beam, we find
\begin{equation} \label{eq:mumubsbar}
    \frac{d \sigma(\mu^+\mu^- \to b \bar s)}{d\cos\theta} = \frac{3}{16} \sigma(\mu^+\mu^- \to b s) \Big(1+\cos^2\theta + \frac{8}{3} A_\text{FB} \cos\theta \Big) ~,
\end{equation}
\begin{equation} \label{eq:mumubbars}
    \frac{d\sigma(\mu^+\mu^- \to \bar b s)}{d\cos\theta}  = \frac{3}{16} \sigma(\mu^+\mu^- \to b s) \Big( 1+\cos^2\theta - \frac{8}{3} A_\text{FB} \cos\theta \Big) ~,
\end{equation}
where $\theta$ is the angle between the $\mu^-$ and the $b$ or $\bar b$, respectively. The above expressions contain the total cross section
\begin{multline}
    \sigma(\mu^+\mu^- \to b s) = \sigma(\mu^+\mu^- \to b \bar s) + \sigma(\mu^+\mu^- \to \bar b s) \\
    = \frac{G_F^2 \alpha^2}{8 \pi^3} |V_{tb} V_{ts}^*|^2 s \left[ (1 - P_+ P_-) \Big( |C_9|^2 + |C_{10}|^2 \Big) - 2 ( P_+ - P_-) \text{Re}(C_9 C_{10}^*) \right] ~.
\end{multline}
as well as the forward backward asymmetry of the $b$-jet, $A_\text{FB}$,
\begin{equation} \label{eq:AFB}
 \frac{4}{3} A_\text{FB} = \frac{( P_+ - P_-) \Big( |C_9|^2 + |C_{10}|^2 \Big) - 2 (1 - P_+ P_-) \text{Re}(C_9 C_{10}^*)}{(1 - P_+ P_-) \Big( |C_9|^2 + |C_{10}|^2 \Big) - 2 ( P_+ - P_-) \text{Re}(C_9 C_{10}^*)} ~.
\end{equation}
As expected from dimensional analysis, the signal cross section grows with the center of mass energy squared, $s$. The forward backward asymmetry is independent of $s$ and predicted to be $A_\text{FB} = +75\%$ if the new physics affects only left-handed muons and $A_\text{FB} = -75\%$ if the new physics affects only right-handed muons.
Note that it enters with a different sign in the $\mu^+\mu^- \to b \bar s$ and $\mu^+\mu^- \to \bar b s$ cross sections. A measurement of $A_\text{FB}$ thus requires charge tagging of the $b$-jets. 
The beam polarizations $P_\mp \in [-1,1]$ specify the fraction of polarized muons and anti-muons respectively, with $P_\mp = +1 (-1)$ indicating purely right-handed (left-handed) beams and the unpolarized limit corresponding to $P_\mp = 0$. A sizable polarization could be a possibility at a muon collider but would likely imply a reduced luminosity~\cite{Ankenbrandt:1999cta}. We find that beam polarization can have important impact on our results.

The Wilson coefficients entering the cross section and forward backward asymmetry should be evaluated at a renormalization scale of the order of the center of mass energy of the muon collider $\mu \sim \sqrt{s}$. We have explicitly checked that renormalization group effects~\cite{Jenkins:2013wua, Alonso:2013hga, Jenkins:2017jig, Jenkins:2017dyc} between the $B$ meson scale and the center of mass of a high energy muon collider are at the few percent level and we therefore neglect them.
\section{Backgrounds and Collider Analysis} \label{sec:backgrounds}

Various background processes to the $\mu^+\mu^- \to b s$ signal need to be considered.
\begin{itemize}
    \item There is an irreducible SM loop contribution to the  $\mu^+\mu^- \to b s$ cross section. At a high energy muon collider, the SM contribution cannot be described by a contact interaction but requires a calculation with dynamical top quarks, $W$ bosons, and $Z$ bosons. We have calculated this cross section for arbitrary $\sqrt{s}$ using \verb|FeynArts|~\cite{Hahn:2000kx} and \verb|FormCalc|~\cite{Hahn:1998yk}. For large center of mass energy, $\sqrt{s} \gg m_t, m_W, m_Z$, we find that the cross section falls with $\sigma_\text{bkgd}^\text{loop} \propto 1/s$ and is completely negligible.
    \item A much more important source of background stems from mistagged di-jet events. We consider $\mu^+ \mu^- \to b \bar b$ events in which one $b$-jet is misidentified as a light jet, as well as $\mu^+ \mu^- \to c \bar c$ and $\mu^+ \mu^- \to q \bar q$ events with light quarks $q = u,d,s$, where one of the charm or light quark jets is identified as a $b$-jet. We analytically calculated the corresponding di-jet cross sections at tree level. We assume that $\mu^+ \mu^- \to t \bar t$ events do not give a relevant background. The corresponding background cross section is therefore
    \begin{equation}
     \sigma_\text{bkgd}^\text{mis-tag} = 2 \sum_{q = u,d,s,c,b} \epsilon_q (1-\epsilon_q)~ \sigma(\mu^+ \mu^- \to q \bar q) ~,
    \end{equation}
    where $\epsilon_b$ is the $b$-tag efficiency and $\epsilon_{u,d,s,c}$ the probabilities that a charm or light quark jet is misidentified as a $b$-jet. 
    For the numerical analysis we follow~\cite{Huang:2021biu} and adopt the values: $\epsilon_b = 70\%$, $\epsilon_c = 10\%$, and $\epsilon_u = \epsilon_d = \epsilon_s = 1\%$. These values are comparable to those that are currently achieved by the ATLAS and CMS experiments at the LHC for jets with transverse momentum up to a few hundred GeV~\cite{CMS:2017wtu, ATLAS:2017bcq}. The performance of traditional flavor taggers decreases significantly for a jet $p_T$ in the multi-TeV regime. However, novel tagging techniques~\cite{PerezCodina:2631478} should improve the performance for very high energy jets to the level quoted above.
    \item Finally, additional backgrounds come from di-jet production through vector boson fusion (in association with forward neutrinos that remain undetected): $\mu^+\mu^- \to b \bar b \nu \bar \nu$, $\mu^+\mu^- \to c \bar c \nu \bar \nu$, or $\mu^+\mu^- \to q \bar q \nu \bar \nu$ with mistagged quarks, and $\mu^+\mu^- \to b \bar s \nu \bar \nu$. These processes are potentially relevant, as the vector boson fusion cross section grows with the center of mass energy~\cite{Costantini:2020stv}. 
    However, this background can be largely removed by cuts on the di-jet invariant mass, which is expected to be $m_{jj} \simeq \sqrt{s}$ for the signal events, but is lower for the background, due to the neutrinos carrying away energy. 
    A di-jet invariant mass resolution of $\sim 2\%$ for 5~TeV di-jets has been achieved at ATLAS~\cite{ATLAS:2017eqx}. We assume that detectors at a future muon collider will perform at least as good. 
    We determine the cross sections of this background using \verb|MadGraph5|~\cite{Alwall:2014hca} and employ a cut on the di-jet invariant mass of $m_{jj}/\sqrt{s} = 1 \pm 0.04$. Such a cut retains $\simeq 95\%$ of the signal but reduces this background to a negligible level. Similarly, also $\mu^+\mu^- \to q \bar q^\prime \mu^+ \mu^-$ and $\mu^+\mu^- \to q \bar q^\prime \mu \nu$ backgrounds are negligible once  the $m_{jj}$ cut is taken into account and we don't consider them here.
\end{itemize}

\begin{figure}[tb]
 \centering
  \includegraphics[width=0.9\textwidth]{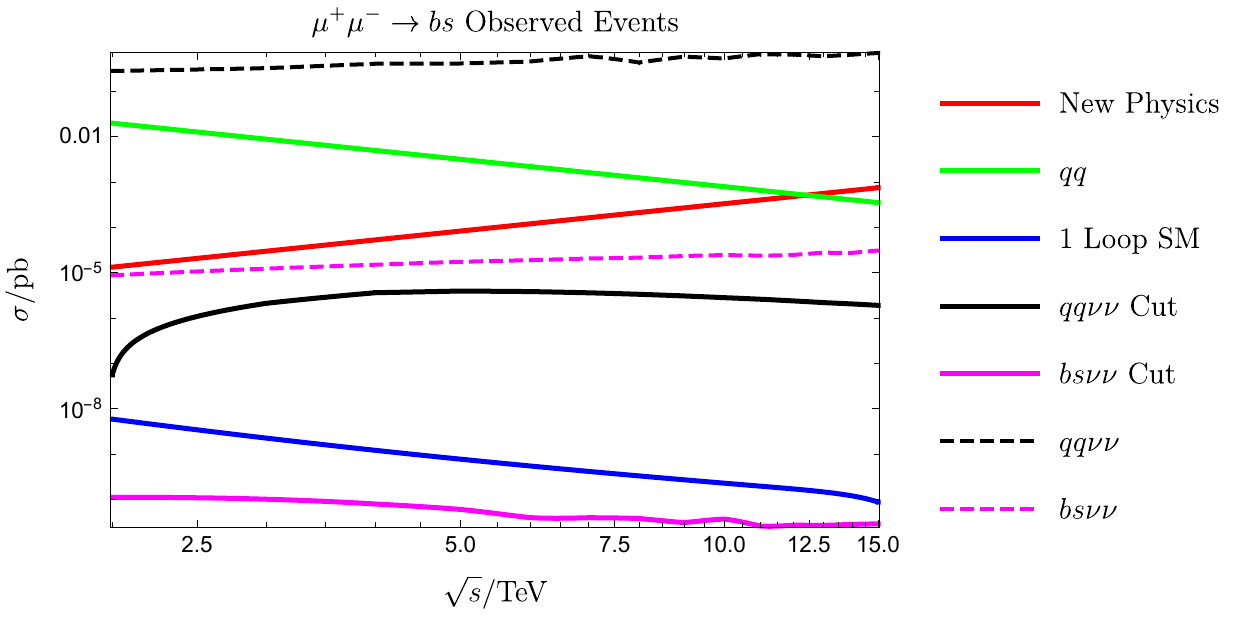} 
  \caption{The cross sections of the $\mu^+\mu^- \to b s$ signal and background processes at a muon collider with center of mass energy $\sqrt{s}$. The shown cross sections take into account $b$-tagging efficiencies and mistag rates. The label $q q$ refers to the sum of $b \bar b$, $c \bar c$ and light quarks. The dashed lines correspond to the vector boson fusion backgrounds without cut on the di-jet invariant mass (see text). The muon beams are assumed to be unpolarized.}
  \label{fig:cross_sections}
\end{figure}

In figure~\ref{fig:cross_sections} we show the cross sections of the new physics $\mu^+\mu^- \to b s$ signal and the background processes mentioned above as a function of the center of mass energy $\sqrt{s}$. In the figure, the muon beams are assumed to be unpolarized. 
For the signal cross section, we assume a new physics benchmark scenario with $C_9 = -0.35$, $C_{10} = +0.35$, well within the $1\sigma$ region determined in~\cite{Altmannshofer:2021qrr}. 
The lines show the cross sections taking into account the $b$-tagging efficiency and mistag rates. The dashed lines show the backgrounds from vector boson fusion without the cut on the di-jet invariant mass.

At low center of mass energy the mistagged di-jet background dominates the signal by orders of magnitude. Among the di-jet backgrounds the $b\bar b$ final state contributes the most, followed by $c \bar c$ and light quarks. The SM loop background and the background from vector boson fusion (with $m_{jj}$ cut) are subdominant and we will neglect them in the following. Signal and background cross sections become comparable for a center of mass energy of around 12~TeV. This suggests that a 10~TeV muon collider should be able to observe the non-standard $\mu^+ \mu^- \to bs$ production with high significance.

\section{Sensitivity Projections} \label{sec:results}

For the sensitivity projections, we focus on a 10~TeV muon collider with integrated luminosity of up to 10~ab$^{-1}$~\cite{muon_forum}. 
At lower center of mass energies (e.g. 6~TeV) we find only very weak sensitivities mainly due to the large $\mu^+ \mu^- \to b \bar b$ background.

\begin{figure}[tb]
 \centering
  \includegraphics[width=0.46\textwidth]{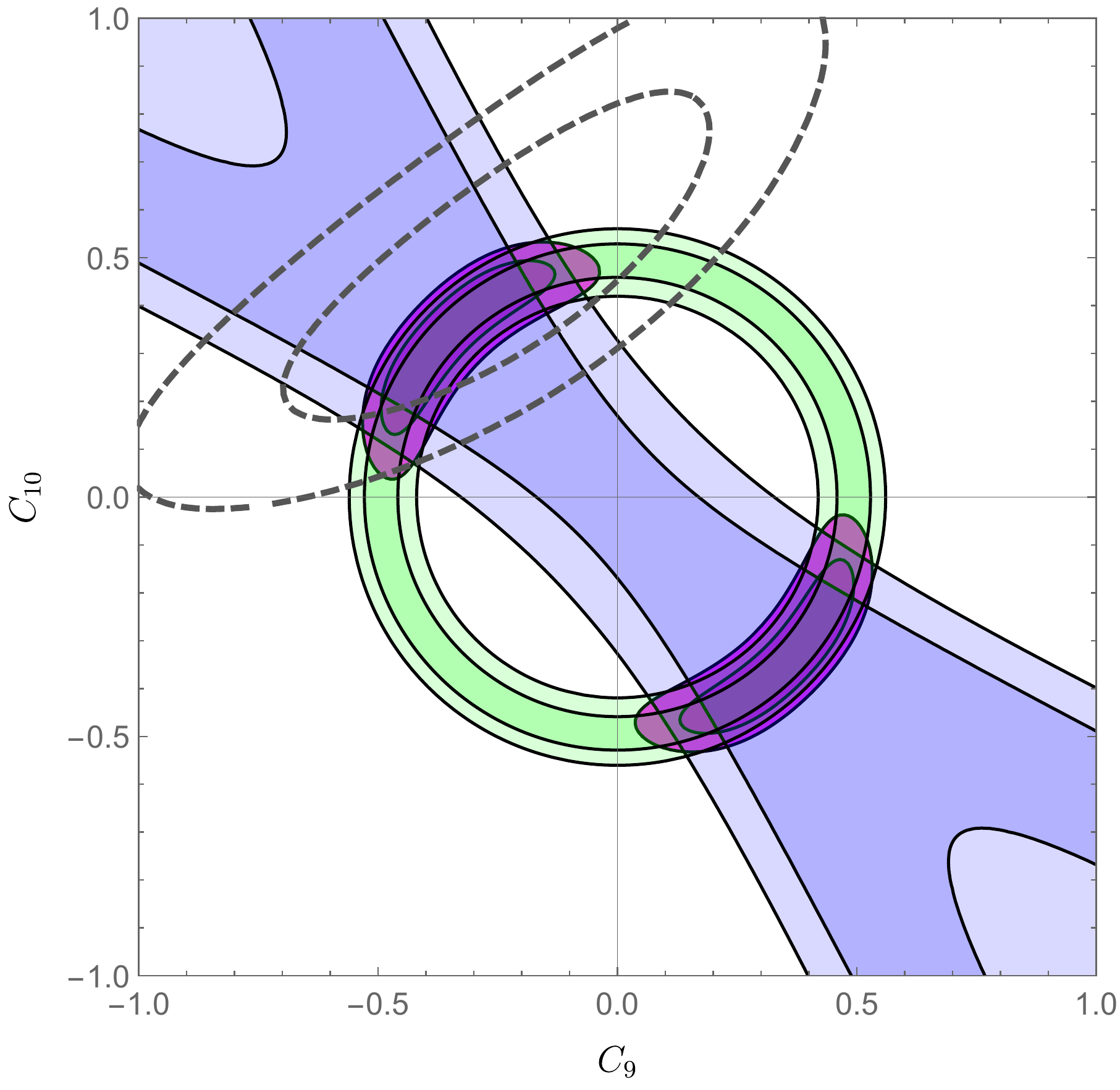} \quad
  \includegraphics[width=0.46\textwidth]{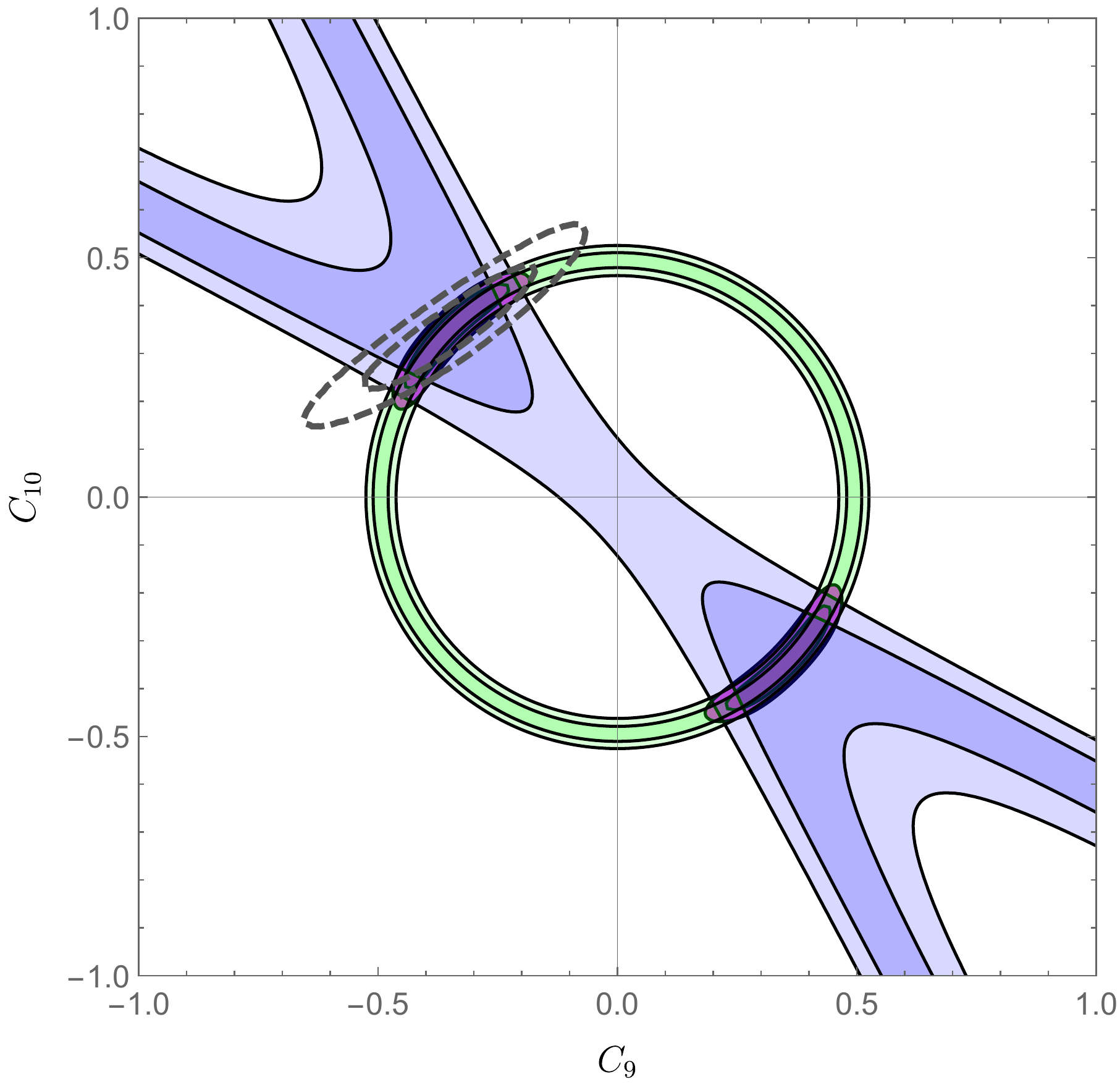}
  \caption{Sensitivity of a 10~TeV muon collider with unpolarized beams in the $C_9$ vs. $C_{10}$ plane assuming a new physics benchmark point with $C_9 = -0.35$, $C_{10} = +0.35$. In green (blue) the region that can be determined by a measurement of the $\mu^+\mu^- \to b s$ cross section (the forward backward asymmetry). The combination is in purple. The left (right) plot assumes an integrated luminosity of 1~ab$^{-1}$ (10~ab$^{-1}$). The dashed black lines are the current best fit region explaining the rare $B$ decay anomalies (left plot) or the expected region after the HL-LHC (right plot).}
  \label{fig:C9_vs_C10_benchmark}
\end{figure}
\begin{figure}[tb]
 \centering
  \includegraphics[width=0.46\textwidth]{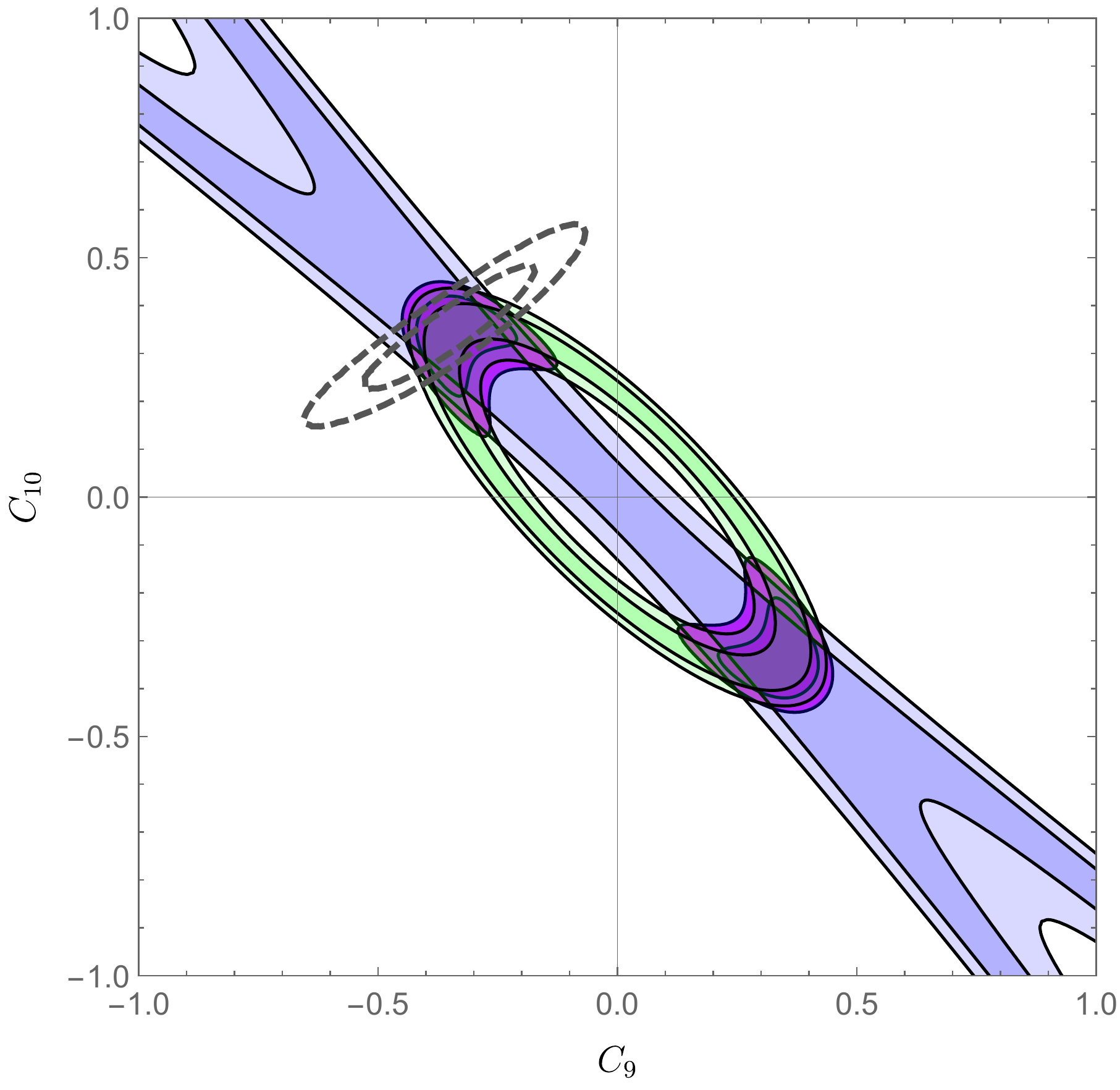}
  \caption{Same as the 10~ab$^{-1}$ plot in figure~\ref{fig:C9_vs_C10_benchmark}, but with a muon beam polarization of $P_- = - P_+ = 50\%$.}
  \label{fig:C9_vs_C10_benchmark_polarized}
\end{figure}

First, we consider a new physics benchmark motivated by an explanation of the current anomalies in the rare $B$ decays, $C_9 = -0.35$, $C_{10} = +0.35$~\cite{Altmannshofer:2021qrr}.
For this benchmark, we determine the number of expected $\mu^+ \mu^- \to b s$ events, taking into account the $b$-tagging efficiencies and mistag rates discussed in section~\ref{sec:backgrounds}. At a center of mass energy of 10~TeV and for unpolarized muon beams, we find the following event numbers
\begin{eqnarray}
    N_\text{total} = {1109 \pm 37} ~,~~~~~  N_\text{bkgd} = {781 \pm 32} ~,~~~ && @~ 1~\text{ab}^{-1} ~, \\
    N_\text{total} = {11089 \pm 188} ~,~~ N_\text{bkgd} = {7809 \pm 179} ~,\,&& @~ 10~\text{ab}^{-1} ~,
\end{eqnarray}
where ``total'' refers to the sum of signal and background events.
For a beam polarization of $P_- = - P_+ = 50\%$ we get 
\begin{eqnarray}
    N_\text{total}= {677 \pm 29} ~,~~~~~  N_\text{bkgd} = {595 \pm 27} ~,~~~ && @~ 1~\text{ab}^{-1} ~, \\
    N_\text{total} = {6767 \pm 145} ~,~~ N_\text{bkgd} = {5947 \pm 142} ~,\,&& @~ 10~\text{ab}^{-1} ~.
\end{eqnarray}
The quoted uncertainties include the statistical as well as a 2\% systematic uncertainty added in quadrature. In all cases the total number of events is significantly above the background prediction. This translates into an expected precision on the measured signal cross section of $\sim 15\%$ at 1~ab$^{-1}$ and $\sim 8\%$ at 10~ab$^{-1}$ for unpolarized beams. For 50\% polarization the precision is considerably lower, $\sim 48\%$ at 1~ab$^{-1}$ and $\sim 25\%$ at 10~ab$^{-1}$. This is expected as the new physics benchmark corresponds to an interaction that only involves left-handed muons. The $1\sigma$ and $2\sigma$ constraints in the $C_9 - C_{10}$ plane from the expected measurement of the cross section are shown in green in figures~\ref{fig:C9_vs_C10_benchmark} and~\ref{fig:C9_vs_C10_benchmark_polarized}. For polarized beams we only show a plot for an integrated luminosity of 10~ab$^{-1}$, as the sensitivity for 1~ab$^{-1}$ is rather poor.

The best precision on the forward backward asymmetry could likely be obtained by performing a likelihood fit to the angular shape in equations~\eqref{eq:mumubsbar} and~\eqref{eq:mumubbars}.
Here, we determine the expected precision by simply splitting the total events into forward and backward categories. As noted in section~\ref{sec:theory}, the measurement of the forward backward asymmetry requires charge tagging of the jets. Denoting the charge tagging efficiency by $\epsilon_\pm$, we have
\begin{eqnarray}
    A_\text{FB}^\text{total} &=& \frac{2\epsilon_\pm - 1}{N_\text{total}} \Big( N_\text{total}^{b, \text{forward}} + N_\text{total}^{\bar b, \text{backward}} - N_\text{total}^{b, \text{backward}} - N_\text{total}^{\bar b, \text{forward}} \Big) \nonumber \\
    &=& (2\epsilon_\pm - 1) \Big( \frac{N_\text{signal}}{N_\text{total}} A_\text{FB} + \frac{N_\text{bkgd}}{N_\text{total}} A_\text{FB}^\text{bkgd} \Big) ~,
\end{eqnarray}
where $A_\text{FB}$ is the forward backward asymmetry of the signal events from equation~\eqref{eq:AFB}.
We use $\epsilon_\pm = 70\%$ which is comparable to the performance achieved at LEP~\cite{DELPHI:2004wzo}.
Taking into account statistical and 2\% systematic uncertainties on the event numbers, we find expected measurements of the total forward backward asymmetry $A_\text{FB}^\text{total} = (25.7 \pm 1.5)\%$ at 1~ab$^{-1}$ and $A_\text{FB}^\text{total} = (25.7 \pm 0.6)\%$ at 10~ab$^{-1}$ for unpolarized beams. The forward backward asymmetry is highly complementary to the cross section and leads to orthogonal constraints in the $C_9 - C_{10}$ plane, see the blue regions in figures~\ref{fig:C9_vs_C10_benchmark} and~\ref{fig:C9_vs_C10_benchmark_polarized}.

The combination of cross section and forward backward asymmetry is shown in purple.
For comparison, the dashed black contours in the plots on the left hand side show the $1\sigma$ and $2\sigma$ best fit region that explains the current rare $B$ decay anomalies. We obtain this region by taking into account the current measurements of the theoretically clean LFU ratios $R_K$ and $R_{K^*}$ from~\cite{LHCb:2017avl, LHCb:2021trn} and the world average of the $B_s \to \mu^+ \mu^-$ branching ratio measurements from~\cite{Altmannshofer:2021qrr}. 
The dashed black contours in the plots on the right hand side show the $1\sigma$ and $2\sigma$ best fit region of the $B$ anomalies that can be expected from LHCb with 300~fb$^{-1}$ based on the sensitivity projections for $R_K$, $R_{K^*}$ and $B_s \to \mu^+ \mu^-$ from~\cite{LHCb:2018roe}.

The plots show that a 10~TeV muon collider can establish a new physics signal with high precision. For unpolarized beamsn already with an integrated luminosity of 1~ab$^{-1}$ the best fit region from the muon collider is much more precise than the region that is currently preferred by the $B$ anomalies. With 10~ab$^{-1}$ the muon collider has a precision that is better than the one expected after the high luminosity LHC.
As there is no interference of the new physics with a SM contribution to the $\mu^+ \mu^- \to b s$ process, the measurements at the muon collider cannot resolve a sign ambiguity in the new physics Wilson coefficients $C_{9,10} \leftrightarrow - C_{9,10}$. 

In the absence of new physics, a 10~TeV muon collider can constrain the size of the Wilson coefficients $C_9$ and $C_{10}$. For unpolarized muon beams we find
\begin{equation}
    \sqrt{|C_9|^2 + |C_{10}|^2} < \begin{cases} 0.25 ~~~ @ ~1~\text{ab}^{-1} \\ 0.17 ~~~ @ ~10~\text{ab}^{-1} \end{cases} ~.
\end{equation}
Already with $1$~ab$^{-1}$, the constraint would be strong enough to exclude a new physics origin of the $B$ anomalies. The constraint on the Wilson coefficients can also be translated into a sensitivity to a high new physics scale. We find
\begin{equation}
    \Lambda_\text{NP} = \left(\frac{4G_F}{\sqrt{2}} |V_{tb} V_{ts}^*| \frac{\alpha}{4\pi}\right)^{-\frac{1}{2}} \big(|C_9|^2 + |C_{10}|^2\big)^{-\frac{1}{4}}  > \begin{cases} 71~\text{TeV} ~~~ @ ~1~\text{ab}^{-1} \\ 86~\text{TeV} ~~~ @ ~10~\text{ab}^{-1} \end{cases} ~.
\end{equation}
This shows that a muon collider has indirect sensitivity to new physics scales far above its center of mass energy and also above the scale of $\sim 35$~TeV that is hinted at by the $B$ anomalies.

\Acknowledgements
We thank Sergo Jindariani, Patrick Meade, and Vladimir Shiltsev for useful correspondence and Logan Morrison for insightful discussions. S.A.G thanks Thomas Hahn and Olivier Mattelaer for useful discussions pertaining to FormCalc, and MadGraph5, respectively. We acknowledge support by the U.S. Department of Energy grant number DE-SC0010107. W.A. acknowledges support by the Munich Institute for Astro- and Particle Physics (MIAPP) which is funded by the Deutsche Forschungsgemeinschaft (DFG, German Research Foundation) under Germany's Excellence Strategy – EXC-2094-390783311.

\bibliographystyle{JHEP}
\bibliography{theBIB}

\end{document}